\documentclass[%
reprint,
superscriptaddress,
groupedaddress,
 amsmath,amssymb,
 aps,
]{revtex4-2}

\usepackage{graphicx}
\usepackage{dcolumn}
\usepackage{bm}
\usepackage{float}
\usepackage{siunitx}
\usepackage{physics}

\usepackage{hyperref}
\hypersetup{
    colorlinks=false,
    citecolor=red,
    linkcolor=red,
    filecolor=red,      
    urlcolor=red,
    }

\usepackage{gensymb}
\usepackage{pgfplots, pgfplotstable}
\pgfplotsset{width=0.45\textwidth,compat=1.9}
\usetikzlibrary{calc}
\usetikzlibrary{positioning}
\pgfmathsetlengthmacro\MajorTickLength{
  \pgfkeysvalueof{/pgfplots/major tick length} * 0.7}
  
\AtBeginDocument{\RenewCommandCopy\qty\SI}

\begin{document}

\title{In-situ-tunable spin-spin interactions in a Penning trap with in-bore optomechanics}

\author{Joseph H. Pham}
\author{Julian Y. Z. Jee}%

\author{Alexander Rischka}
\altaffiliation[Current address: ]{Q-CTRL Pty Ltd, Sydney, NSW 2000, Australia}
\author{Michael J. Biercuk}%
\author{Robert N. Wolf}
\email[e-mail: ]{robert.wolf@sydney.edu.au}
\affiliation{%
ARC Centre for Engineered Quantum Systems, School of Physics, The University of Sydney, Sydney, NSW 2006, Australia
}%

\date{\today}

\begin{abstract}

Experimental implementations of quantum simulation must balance the controllability of the quantum system under test with decoherence typically introduced through interaction with external control fields. The ratio of coherent interaction strength to decoherence induced by stimulated emission in atomic systems is typically determined by hardware constraints, limiting the flexibility needed to explore different operating regimes. Here, we present an optomechanical system for in-situ tuning of the coherent spin-motion and spin-spin interaction strength in two-dimensional ion crystals confined in a Penning trap. The system introduces active optical positioners into the tightly constrained space of the bore of a superconducting magnet, allowing adjustability of the key hardware parameter which determines the ratio of coherent to incoherent light-matter interaction for fixed optical power. Using precision closed-loop piezo-actuated positioners, the system permits in-situ tuning of the angle-of-incidence of laser beams incident on the ion crystal up to $\theta_{\mathrm{ODF}}\approx 28^\circ$. We characterize the system using measurements of the induced mean-field spin precession under the application of an optical dipole force in ion crystals cooled below the Doppler limit through electromagnetically induced transparency cooling. These experiments show approximately a $\times2$ variation in the ratio of the coherent to incoherent interaction strength with changing $\theta_{\text{ODF}}$, consistent with theoretical predictions.  We characterize system stability over 6000 seconds; rigid mounting of optomechanics to the ion trap structure reduces differential laser movements to approximately $2\times 10^{-3}$ degrees per hour, enabling long-duration experiments. These technical developments will be crucial in future quantum simulations and sensing applications.

\end{abstract}

\maketitle

\section{Introduction}

Employing quantum systems to solve computationally complex problems is an area of intense research \cite{Altman2021, Alexeev2021} due to the potential to deliver solutions faster than, or with superior outputs relative to classical computers. One promising approach towards building quantum computers and simulators uses ion traps as the underlying technology to store and manipulate quantum information carriers~\cite{Bermudez2017, Monroe2021, Marciniak2022}. Ion traps use electromagnetic fields to confine charged particles in an ultra-high-vacuum environment, offering access to long-lived quantum systems with straightforward controllability via optical and microwave fields. Penning traps, in particular, can confine hundreds of ions in stable naturally ordered arrays via a combination of a strong homogeneous magnetic field and a quadrupole electric field and enable coherent optical manipulation of the spin and motional degrees of freedom of the ions \cite{Britton2012, Ball2019}. The large number of accessible ions -- and, as a consequence, qubits -- makes the Penning trap an attractive candidate for quantum simulation of strongly correlated spin systems predicted to govern, for example, magnetic frustration and high-temperature superconductors \cite{G_rttner_2017, Shankar2022a}. 

In trapped-ion systems, lasers are used to coherently induce a coupling between the internal ion-qubit states and their external motional states, for example, using Raman transitions \cite{Lee2005}. To achieve these couplings, a light-shift gate can be used to implement a spin-dependent optical dipole force (ODF) \cite{Leibfried2003, Biercuk2011, Carter2023}, however, for their practical implementation to be efficient, the coherent and incoherent effects of the applied light-matter interaction must be carefully balanced. The magnitude of the ODF, $F_0$, determines the spin-motion and spin-spin coupling rate; an increase will lead to shorter required interaction times, which is advantageous when devices are limited by qubit decoherence. However, increasing $F_0$ by using higher optical power of the ODF lasers also proportionally increases the off-resonant scattering rate, $\Gamma$, which is typically the dominant source of decoherence in these systems \cite{Britton2012, Carter2023, Uys2010}. 

Here, we present a technique to improve the ratio of coherent to incoherent interaction strength, $F_0/\Gamma$, by introducing in-situ tunable ODF laser beams through an active in-bore optomechanical system.  This approach allows tuning of the critical parameter -- the laser separation angle $\theta_\mathrm{ODF}$ -- over a range of about $\qty{12}{\degree}$ to $\qty{28}{\degree}$. By using remotely controllable piezo-actuated optomechanics inside the bore of the superconducting magnet, we can control the Raman laser beam positions and directions with high precision and repeatability. Similarly, this setup is also used to deliver additional lasers enabling electromagnetically induced transparency (EIT) cooling of large two-dimensional ion crystals. In combination, these two techniques facilitate favorable coherent to incoherent interaction strengths while allowing us to remain in the Lamb-Dicke regime for low effective ODF wavelengths, generated by large ODF beam separation angles. We experimentally characterize this system by measuring the mean-field-induced spin precession for sub-Doppler cooled ion crystals under applications of the ODF; the coherent to incoherent interaction strength increases linearly with the beam opening angle, but this increase is suppressed in the absence of EIT cooling. We probe the mechanical stability of this system using an interferometer and directly with ions, arriving at root-mean-square Raman beat-note phase fluctuations of about $\qty{7}{\degree}$ and angular alignment drifts on the order of $0.002\degree/\mathrm{h}$ for the highest opening angle of $\theta_\mathrm{ODF}\approx\qty{28}{\degree}$.

The remainder of the manuscript is structured as follows. Section \ref{sec.exp_design} describes the experimental design and implementation in detail. Section \ref{sec.exp_char} presents the characterization of the system using ODF with Doppler and sub-Doppler cooled ion crystals. Finally, section \ref{sec.sum} provides a summary and outlook.

\section{Adjustable laser delivery system for in-situ-tunable coherent interactions}
\label{sec.exp_design}

In the following, we present the experimental design and implementation of the optomechanical system used for in-situ Raman-beam angular control. In section \ref{subsec.PTsetup}, we begin by describing the Penning trap apparatus, which brings severe spatial-access limitations due to the bore of the superconducting magnet, previously described in detail in Ref. \cite{Ball2019}. We introduce the theoretical foundations for the role of the Raman beam angle in determining the coherent interaction strength in \ref{subsec.OpPrinciple}, and introduce our active in-bore optomechanical solution in section \ref{subsec.InBoreOpt}. We also discuss the Raman laser beam preparation and delivery system mounted on the ion trap setup in section \ref{subsec.LaserSys}.

\subsection{Penning-trap setup}
\label{subsec.PTsetup}

\begin{figure*}
\input{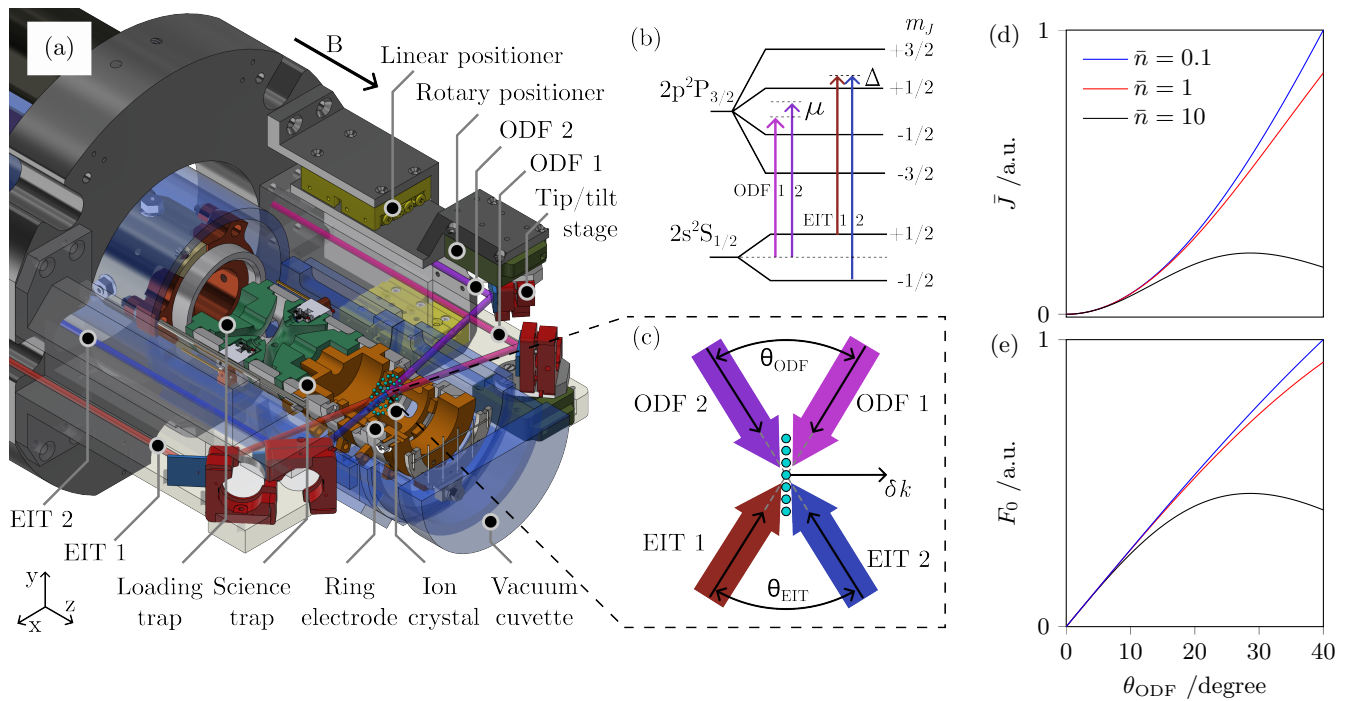}
\caption{(a) Sectional view of the in-bore optomechanical system to position ODF and EIT beams at the position of the ion crystal. The angle $\theta_\mathrm{ODF}$ between both ODF beams can be varied by moving mirrors mounted on tip/tilt stages via rotary and linear positioners. The EIT beams are aligned using only tip/tilt stages. Some parts are made transparent for visibility. (b) Simplified $^9\mathrm{Be}^+$ energy level scheme showing only the relevant ODF and EIT transitions near $\qty{313}{\nano\meter}$. $\mu$ is the frequency difference between the two ODF beams, $\Delta$ is the frequency detuning of the EIT beams from the excited state. (c) Detailed view of the beam geometry at the center of the trap. The difference wave vector $\delta k$ of the ODF and EIT beams is ideally aligned perpendicular to the plane of the 2D ion crystal. (d) Qualitative dependence of the spin-spin coupling coefficient $\bar{J}$ and (e) force $F_0$ with respect to the ODF beam angle and for different mean occupation numbers $\bar{n}$ of the center-of-mass mode. In our system, $\bar{n}\approx 10$ corresponds to the Doppler limit, $\bar{n}\approx 1$ is the occupation after EIT cooling, and $\bar{n}\approx 0.1$ might be in reach with increased EIT cooling performance. The difference between Doppler and EIT cooling becomes significant for beam angles above $20\degree$.}
\label{fig.F0_CAD}
\end{figure*}

We create and confine Beryllium-9 ions in a Penning trap system consisting of two dedicated trapping regions for ion loading (loading trap) and experiments requiring high optical access (science trap), see Fig. \ref{fig.F0_CAD}. The latter has an inner diameter of $20\,\mathrm{mm}$ and radial ($xy$-plane) access for laser and millimetre wave radiation as well as fluorescence photon collection provided by eight azimuthally distributed openings in the central ring electrode. Relevant optical transitions (Fig. \ref{fig.F0_CAD}b) are all in the ultraviolet (UV), with wavelengths of approximately $313\,\mathrm{nm}$; accordingly, the electrode structure is housed in a quartz vacuum cuvette with eight $313\,\mathrm{nm}$ anti-reflection coated windows aligned to the radial trap openings. The vacuum cuvette has an outer diameter of $68\,\mathrm{mm}$. 

The whole ion trap in its vacuum system is mounted on a non-magnetic hexapod and inserted into the $150\,\mathrm{mm}$-diameter and $839\,\mathrm{mm}$-long bore of a $2\,\mathrm{T}$ superconducting magnet. This leaves a $41\,\mathrm{mm}$ wide annular opening between the vacuum system and the bore of the magnet to accommodate all control-field routing, steering, and tuning. In particular, we must use this region for radial Doppler laser cooling \cite{Itano1988}, $\sim 55\,\mathrm{GHz}$ millimeter wave delivery via a rigid waveguide for global spin control, optical imaging of the ion crystal in the radial direction \cite{Ball2019, Mitchell1998a}, and atomic and ionic fluorescence photon counting using a photon multiplier.  

It is in this shared and highly congested region that we must also accommodate the two pairs of Raman lasers which generate a spin-dependent ODF and enable EIT cooling.  Further, these Raman lasers must be routed to be incident on the ion crystal at an angle to the trap axis, symmetric about the ion plane (Fig.~\ref{fig.F0_CAD}c).  Accommodating this beam configuration within the tight annular region between the trap and the inner diameter of the magnet bore is what we must address through a novel laser-delivery system design.

\subsection{Operating principle for tunable coherent interactions}
\label{subsec.OpPrinciple}

The two pairs of Raman laser beams, used to generate a spin-dependent ODF and for EIT cooling, must produce a difference wave vector, $\delta \vec{k}$, aligned with the targeted motional direction, typically the axial ($\Vec{z}$) direction. To achieve this, the lasers cross at the position of the ion crystal under an angle $\theta_\mathrm{ODF}$ and $\theta_\mathrm{EIT}$ (Fig.~\ref{fig.F0_CAD}c). 
For the ODF, 
\begin{equation}
\delta k = \lvert \vec{k}_1 - \vec{k}_2 \rvert \approx (\lvert k_1 + k_2 \rvert)\sin(\theta_\mathrm{ODF}/2), 
\label{eq.delta_k}
\end{equation}
where $\vec{k}_{1,2}$ are the wave-vectors of the two ODF lasers. The Hamiltonian for the ODF can be expressed as \cite{Britton2012}
\begin{equation}
\hat{H}_\mathrm{ODF} = 
F_0 \cos (\mu t)\sum_{i=1}^{N}\hat{z}_i\hat{\sigma}^z_i,
\end{equation}
where $N$ is the number of ions, $\mu$ is the frequency difference of the two ODF beams, and $\hat{\sigma}^z_i$ is the Pauli-$z$ spin operator of ion $i$. Here, it is assumed that the axial extent of the ion motion, $\hat{z_i}$, is small compared to the ODF effective wavelength, $\lambda_\mathrm{ODF}=2\pi/\delta k$. 

The magnitude of the resulting ODF, $F_0$, is given by \cite{Bohnet_2016}
\begin{equation}
F_0 = \hbar\lvert\delta_\mathrm{AC}\rvert\delta k \exp(-\delta k^2 \langle \hat{z}_i^2\rangle/2),
\label{eq.F0}
\end{equation}
where $\lvert\delta_\mathrm{AC}\rvert$ depends on the strength of the ODFs' electric field and $\exp(-\delta k^2 \langle \hat{z}_i^2\rangle/2)$ is the Debye-Waller factor accounting for the departure from the Lamb-Dicke regime \cite{wineland1997}. The dependence on the ions' effective temperature $\bar{n}$ arises from the relationship $\langle\hat{z}_i^2\rangle\simeq \frac{\hbar}{2M\omega_{\mathrm{COM}}}(2\bar{n}+1)$, where $\langle\hat{z}_i^2\rangle$ is the average root mean square axial extent of the ions with mass $M$ and center-of-mass (COM) mode frequency $\omega_\mathrm{COM}/2\pi$.

Equation (\ref{eq.F0}) implies that a stronger optical dipole force, $F_0$, can be achieved for fixed optical power, $\lvert\delta_\mathrm{AC}\rvert$, by increasing the difference wave vector, $\delta k$, and reducing the axial extent of the ion, $\hat{z}_i$. Similarly, the pairwise spin-spin interaction coefficients, $J_{i,j}$, between ions $i,j$, as shown later, may scale quadratically with $\delta k$. In particular, we focus on the dependence of $\delta k$ on $\theta_\mathrm{ODF}$, given in  equations ($\ref{eq.delta_k}$) and ($\ref{eq.F0}$).  Tuning $\theta_\mathrm{ODF}$ allows an enhancement of $F_0$, while the total incoherent emission rate, $\Gamma = (\Gamma_\mathrm{Ram}+\Gamma_\mathrm{el})/2$, which is composed of both Raman and elastic Rayleigh scattering, is only little affected. 

The above discussion, in combination with Raman-beam-polarisation requirements to null the AC Stark shift of individual beams \cite{Britton2012}, excludes certain beam geometries and motivates the idea of a radial beam delivery at an angle to the ion plane. Figures \ref{fig.F0_CAD}d,e illustrate the qualitative dependence of $\bar{J}\approx J_{i,j} $ and $F_0$ on $\theta_\mathrm{ODF}$ for different effective ion temperatures, respectively. The direct link between the achievable $F_0$ and the value of $\theta_\mathrm{ODF}$, evident through these theoretical data, motivates the design of the optomechanical system described next.

\subsection{In-bore optomechanics and beam-delivery system}
\label{subsec.InBoreOpt}

\begin{figure*}[h!tb]
     \centering
         \includegraphics[width=0.95\textwidth]{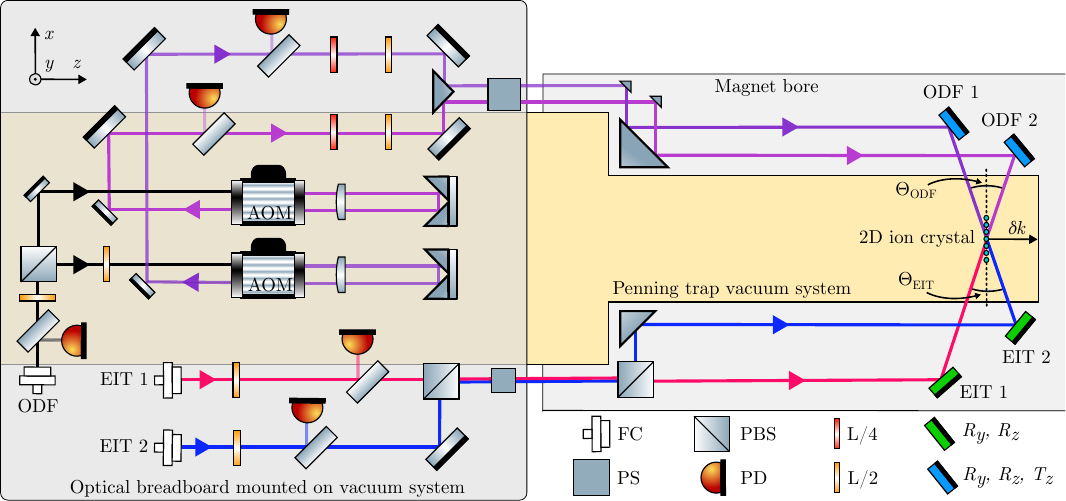}
     \caption{Simplified schematic illustration of the optical system for ODF and EIT beam delivery. All UV light is delivered via UV-compatible fibers to an optical breadboard mounted on the Penning trap vacuum system outside the magnet's bore. After the ODF fiber coupler (FC), a beam sampler deflects a small part of the light to a photodiode for power stabilisation via a PID loop (not shown). The main beam is split into two beams via a polarising beam splitter (PBS) and a half-wave plate (L/2) to balance their power. Each beam is sent through a double-pass acousto-optic modulator (AOM) to generate a frequency difference $\mu$ between both beams. Next, they are sent through beam samplers to monitor a part of their power on a photodiode (PD), followed by a half-wave and quarter-wave plate (L/4) to adjust their polarisation such that each beam does not generate a differential AC Stark-shift at the ion crystal. Both beams are sent via a knife-edge prism into a closely-spaced parallel beam path through a periscope (PS) into the magnet's bore. A set of fixed right-angle prism mirrors deflects the beams towards the piezo-actuated in-bore mirrors, which allow rotation around the $y$ and $z$-axis and a translation along $z$-axis to generate different ODF beam separation angles $\theta_\mathrm{ODF}$. Both ODF beams cross at the center of the ion crystals, and their difference wave vector $\delta k_\mathrm{ODF}$ can be aligned to the direction of the axial modes of the 2D ion crystal. Since both EIT beams are frequency-doubled in separate second-harmonic generation (SHG) units, they arrive in separate fibers at the optical breadboard. Each EIT beam is sent through a half-wave plate to adjust its polarization (EIT 1 $p$-polarized for $\pi$ transition, EIT 2 $s$-polarized for $\sigma^+$ transition), and a beam sampler deflects a small part of the light to a photodiode for power stabilisation via a PID loop (not shown). Due to their orthogonal polarizations, both beams are co-aligned on a PBS and sent through a periscope into the magnet's bore. There, another PBS separates them again until they are deflected into the trap via piezo-actuated mirrors, which allow a tip/tilt beam-angle adjustment.}
    \label{fig:trap_optics}
\end{figure*}

Pursuing radial beam delivery with sufficient tunability to adjust $\theta_\mathrm{ODF}$ while ensuring precision alignment of the interference wavefronts with the ion crystal presents a major optical-system-design challenge.  The inaccessibility of optical elements inside the bore means manual kinematic optomechanics can only be operated outside the bore, $\geq 500\,\mathrm{mm}$ away from the ion crystal. At these distances, small angular changes lead to beam position changes, which substantially alter laser-ion spatial overlap.  Consequently, for optimal performance and minimal disruption, it is preferable to make angular adjustments as close as possible to the ion crystal. 

To achieve this capability, we developed a beam delivery system using non-magnetic, piezo-actuated positioners inside the magnet bore, close to the ion crystal. As shown in Fig. \ref{fig.F0_CAD}a, the two ODF mirrors (Thorlabs rectangular UV-enhanced aluminium, custom) are each mounted on a clear-edge mirror mount (SmarAct STT-12.7) providing tip ($R_z$) and tilt ($R_y$) angular adjustments with a resolution of about $\qty{0.0001}{\degree}$ in open-loop operation. These mounts are, in turn, mounted on a rotary positioner with an encoder (SmarAct SR-2013) providing $100\,\degree$ rotation and angular repeatability of $\delta R_y =0.0014\,\mathrm{\degree}$ in closed-loop operation. Finally, the combined system is mounted on a linear positioner with encoder (SmarAct SLC-1730), providing travel of $21\,\mathrm{mm}$ and repeatability of $\delta T_z=30\,\mathrm{nm}$ in closed-loop operation. 

This system allows for a variation of the ODF beam separation angle $\theta_\mathrm{ODF}$ between $12\,\degree$ and $36\,\degree$; the lower and upper limits are the result of geometrical constraints due to the mirror mounts and the trap-electrode openings, respectively. Closed-loop operation in $R_y$ and $T_z$ allows adjustment of the ODF beam angle $\theta_\mathrm{ODF}$ with millidegree reproducibility. Sub-millidegree adjustments can be achieved in both directions, $R_y$ and $R_z$, using the open-loop actuated mirror mount. 

Different from the ODF mirrors, the EIT mirrors (Thorlabs rectangular UV-enhanced aluminium, custom) are only mounted on tip/tilt clear-edge mirror mounts (SmarAct STT-12.7), without additional rotational or translational positioners since the EIT beam separation angle of $\theta_\mathrm{EIT}\approx 18\,\degree$ can be fixed. The tip/tilt stage allows $\pm2\,\degree$ to align the effective wave vector $\delta k_\mathrm{EIT}$ to the crystal rotation axis. 

\begin{figure}[h!tb]
     \centering
         \includegraphics[width=0.45\textwidth]{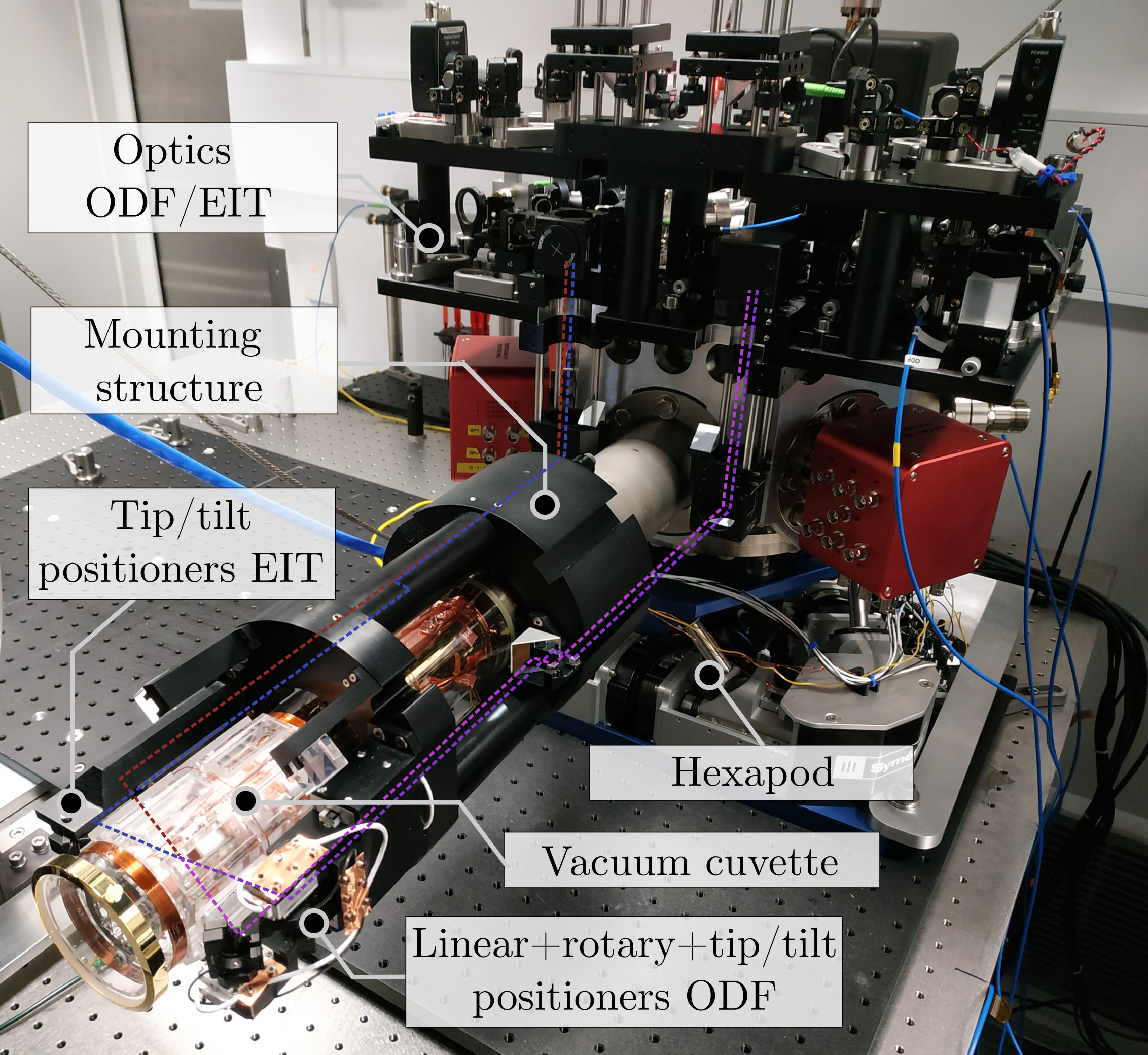}
     \caption{Photograph of the Penning trap setup with mounted in-bore optomechanics (front) and laser beam preparation setup mounted on the vacuum system (back). The whole setup is installed on a hexapod for alignment to the magnetic field. The ODF and EIT beam paths are indicated, similar to Figures \ref{fig.F0_CAD} and \ref{fig:trap_optics}.}
    \label{fig:trap_photo}
\end{figure}

Lasers propagate free space inside the magnet bore from fiber couplers mounted adjacent to the entrance of the magnet bore (Figs. \ref{fig:trap_optics}, \ref{fig:trap_photo}). To mitigate beam pointing instabilities due to differential movement between the ion crystal and UV optics preparing the ODF and EIT beams, an optomechanical breadboard is mounted on the Penning trap vacuum system right before the entrance of the bore of the superconducting magnet. Here, a single UV-compatible fiber delivers the light for the ODF, which is split into two beams by a polarising beamsplitter and a half-wave plate to balance their power. Both beams are then sent through double-pass AOMs (Gooch \& Housego I-M110-3C10BB-3-GH27), which together generate a frequency difference $\mu/2\pi$ near the motional mode frequency, e.g., $\omega_\mathrm{COM}/2\pi \approx \qty{1.1}{\mega\hertz}$. Next, each beam is sent through half-wave and quarter-wave plates to adjust its polarisation such that both create a spin-dependent ODF of the same strength but opposite sign without generating individual-beam AC Stark shifts. 

The ODF beams then enter the magnet bore and propagate parallel to the trap axis before being reflected from the adjustable mirrors described above.  After passing through the vacuum envelope and the ion crystal, the ODF beams leave the trap volume through identical openings on the opposite side of the center ring electrode, after which they are reflected by the EIT mirrors into a beam dump area.

The EIT beams are delivered from the same optical breadboard mounted on the Penning trap vacuum system. Both beams are sent through half-wave plates to adjust their polarisation such that they can be co-aligned on a polarising beam splitter (acting as a beam combiner) before entering the bore. Further in the bore, the beams are spatially separated again by another polarizing beam splitter before they are reflected towards the ion crystal. This co-aligned travel reduces the effect of environment-induced differential phase fluctuations, such as due to temperature, pressure, and humidity differences.

\subsection{Laser sources}
\label{subsec.LaserSys}

In general, the laser beams to generate the ODF and for EIT cooling are produced via sum frequency generation (SFG) of infrared lasers near $1051\,\mathrm{nm}$ and $1550\,\mathrm{nm}$ to generate light near $626\,\mathrm{nm}$, followed by second-harmonic generation (SHG) to $313\,\mathrm{nm}$ \cite{Wilson2011}. For the ODF, both beams are only separated by the motional mode frequency, which can be achieved conveniently using AOMs as described before and shown in Fig. \ref{fig:trap_optics}. Given that the qubit is represented by the Zeeman-splitted ground states of the valance electron spin in $^9\text{Be}^+$, specifically within the 2s$^2$S$_{1/2}$ manifold as $\ket{\uparrow} = \ket{m_j = + 1/2}$ and $\ket{\downarrow} = \ket{m_j = - 1/2}$, see Fig. \ref{fig.F0_CAD}b, both laser beams must bridge this ground-state Zeeman splitting with $\omega_\mathrm{SF}/2\pi \approx 55\,\mathrm{GHz}$ at $2\,\mathrm{T}$ magnetic field for effective EIT cooling and be phase coherent at the ion crystal. 

 Our approach to generating phase-coherent lasers separated by tens to hundreds of gigahertz uses a single infrared laser near $1550\,\mathrm{nm}$ to which we apply double-sideband suppressed-carrier amplitude modulation at $(\omega_\mathrm{SF}/2\pi)/4 \approx13.75\,\mathrm{GHz}$ via an electro-optic modulator (iXblue MXER-LN-20-PD-P-P-FA-FA-30dB). This results in two co-propagating beams near $626\,\mathrm{nm}$ with a frequency difference of $(\omega_\mathrm{SF}/2\pi)/2 \approx 27.5\,\mathrm{GHz}$ after SFG. A Mach-Zehnder beam splitter spatially separates these two beams \cite{Marciniak2020}, which can subsequently be sent to individual SHG units. This strategy mitigates the challenges associated with phase-locking separate lasers at such large frequency differences~\cite{Jordan2019, Mielke2021a}. The generated $313\,\mathrm{nm}$ beams are sent via UV-compatible fibers \cite{Marciniak2017} to the optical setup described above.

\section{Experimental Characterization}
\label{sec.exp_char}

To test the effectiveness of our beam delivery system, we conducted experiments on the cooling efficiency of our EIT cooling scheme. We achieve this by investigating the spin decoherence in the centre-of-mass mode through measurements of ODF-induced spin-motional entanglement. Subsequently, we characterize the dominant decoherence source in our experiment, ODF-induced off-resonant scattering, by observing decoherence at a frequency far detuned from any motional modes \cite{Bohnet_2016}. These measurements, which are explained in section \ref{subsec.EITExp}, provide essential parameters for extracting $F_0$, as shown in section \ref{subsec.F0exp}. Additionally, in section \ref{subsec.StabExp}, we address the central concern of beam-path stability by employing an experimental procedure that focuses on assessing beam angles with respect to the ion crystal and path length differences with an interferometer. 

\subsection{Characterization of EIT cooling efficiency}
\label{subsec.EITExp}

\begin{figure}[t]
\input{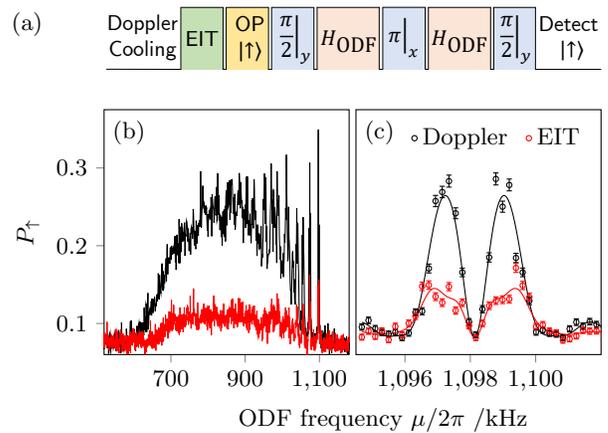}
\caption{a) Experimental pulse sequence for a spin-echo experiment to determine the temperature of the axial modes. `EIT' is the optional EIT cooling period, `OP' is the optical pumping to $\ket{\uparrow}$, `$H_\mathrm{ODF}$' is the application of the ODF beams for time $\tau$.
b) ODF frequency scan over the drumhead modes of a 2D crystal with 125 ions, with (red) and without (black) EIT cooling. 
c) Temperature measurement of the COM mode near $\omega_\mathrm{COM}/2\pi=\qty{1098}{\kilo\hertz}$. Symbols represent the measured data, while solid lines are fits of Eq. (\ref{eq.COM_mode}) to the data with only the COM frequency and the occupation number as free parameters, resulting in $\bar{n}_\mathrm{COM}=10.7(5)$ using $\qty{10}{\milli\second}$ of Doppler cooling, and $\bar{n}_\mathrm{COM}=1.27(20)$ when followed by $t_\mathrm{EIT}=\qty{850}{\mu\second}$ of EIT cooling.  EIT laser parameters are \qty{4}{\milli\watt} power, detuned from the excited state by \qty{350}{MHz}, and $1/e^2$ beam diameters of about \qty{1.1}{\milli\meter}. The error bars represent the statistical uncertainty of the number of photon counts.}
\label{fig.drumhead_modes}
\end{figure}

We characterise the effectiveness of our EIT cooling setup for the axial modes of 2D ion crystals through motional-mode thermometry \cite{Sawyer2012}. The experimental sequence, see Fig. \ref{fig.drumhead_modes}a, begins with state preparation, consisting of Doppler laser cooling, followed by optional EIT cooling for a duration of $t_\mathrm{EIT}$, optical pumping to $\ket{\uparrow}$, and a global $(\pi/2)_y$ resonant mmW pulse to prepare the ions in a spin superposition state. The ODF interaction, which causes motion-dependent phase accumulation, is then applied for $2\tau$, interleaved by a spin echo $(\pi)_x$ pulse to suppress magnetic-field-drift-induced decoherence. A final $(\pi/2)_y$ pulse is applied to map the dephasing to the population in the $\ket{\uparrow}$ spin state, for detection in a spin-selective readout procedure. In this protocol, measurement of spin population $P_{\uparrow}$ is a direct proxy for phonon occupancy of associated motional modes.

In our experimental characterization, we vary the ODF frequency $\mu$ in each iteration and measure the bright-state population, $P_{\uparrow}$. Figure \ref{fig.drumhead_modes}b shows the measured bright-state fraction as a function of $\mu$, scanned over all axial motional modes. As per previous demonstrations~\cite{Sawyer2012}, we observe a large unresolved region of modes, potentially due to elevated temperature of the $\vec{E}\times\vec{B}$ modes \cite{Shankar2020b, Johnson2023}, in addition to well-defined signals for the three highest-frequency modes near the center-of-mass mode frequency, $\omega_{\textrm{COM}}/2\pi\approx \qty{1.1}{\mega\hertz}$. In the presence of EIT cooling, the strength of all signals is decreased substantially, and the higher-frequency modes remain clearly defined. 

We determine the mean occupation number $\bar{n}_\mathrm{COM}$ of the COM mode by fitting an analytical expression to measured data in Fig.~\ref{fig.drumhead_modes}c~\cite{Sawyer2012, Jordan2019},  
\begin{equation}
    P_{\uparrow} = \frac{1}{2} \left( 1-\exp(-\Gamma2\tau) C_\mathrm{ss} C_\mathrm{sm}\right),
    \label{eq.COM_mode}
\end{equation}
where $C_\mathrm{ss} = (\cos(4J))^{N-1}$ describes the phonon-mediated spin-spin interaction and $C_\mathrm{sm} =\exp(-2\left| \alpha \right|^2 (2\bar{n}_\mathrm{COM}+1))$ describes the dephasing due to spin-motion coupling. Here, $J$ is the spin-spin coupling coefficient, $\alpha$ is the spin-dependent displacement amplitude \cite{Sawyer2012}, $N$ is the number of ions, and $\tau$ is the duration of the ODF interaction in one arm of the spin-echo sequence. Both $J$ and $\alpha$ depend on the force, $F_0$, which is determined as described in the next section. 

We determine the number of ions, $N$, by analyzing the single-layer to bilayer transition crystal rotation frequency using ion-crystal conformation imaging perpendicular to its rotation axis~\cite{Biercuk2010}. We independently determine the off-resonant scattering rate, $\Gamma$, by measuring $P_{\uparrow}(\tau)$ at large detunings $\delta=\mu-\omega_\mathrm{COM}$ from the COM mode, such that $C_\mathrm{sm}\approx C_\mathrm{ss} \approx 1$, which gives a direct probe of decoherence. 

The fit in Fig.~\ref{fig.drumhead_modes}c considers the COM mode frequency $\omega_\mathrm{COM}/2\pi$ and the mean occupation number $\bar{n}_\mathrm{COM}$ as the free fit parameters. By applying EIT cooling for \qty{850}{\micro\second}, we can reduce the mean occupation number of the COM mode by about a factor of eight from $\bar{n}_\mathrm{COM}\approx 11$ at the Doppler-cooling limit to $\bar{n}_\mathrm{COM}\approx 1.3$.

\subsection{Quantifying the adjustable coherent-to-incoherent coupling strength}
\label{subsec.F0exp}

\begin{figure}[h!t]%
\input{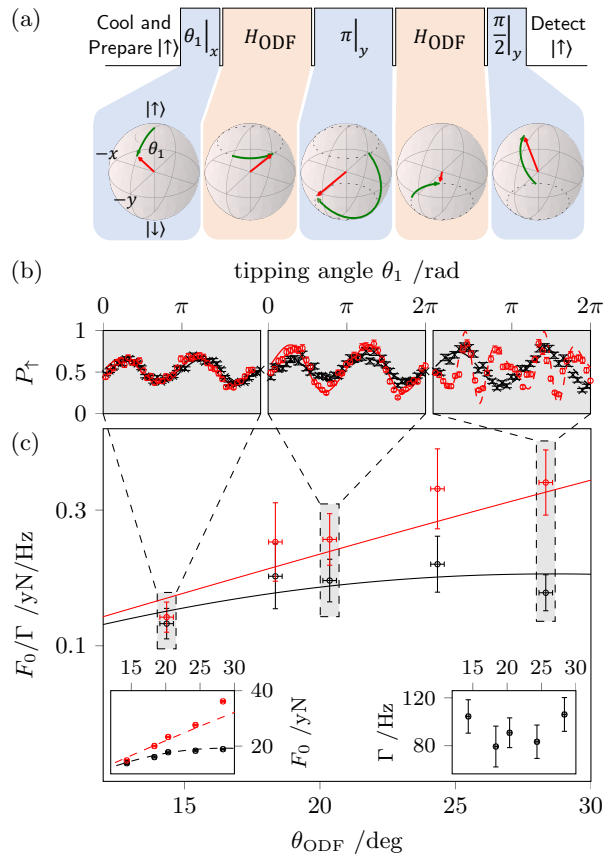}
\caption{a) Experimental pulse sequence to determine the uniform spin-spin coupling strength $\bar{J}$ from which the force $F_0$ can be determined. The sequence is identical to Fig. \ref{fig.drumhead_modes}a, apart from the first mmW tipping pulse whose angle $\theta_1$ is varied, and the middle $\pi$-pulse, which is rotated around the $y$-axis in the Bloch-sphere representation. The green arc shows the rotation, while the red arrow shows the spin state on the Bloch sphere at the end of the interaction.
b) Bright-state population at the end of the pulse sequence with (red) and without (black) additional EIT cooling. The symbols are data taken at $\theta_{\mathrm{ODF}} \approx \qty{14}{\degree}$ (left), $\qty{20}{\degree}$ (middle) and $\qty{28}{\degree}$ (right) at $\tau = \qty{500}{\micro\s}$ and $\delta/2\pi=\qty{2}{\kilo\hertz}$. The lines are fitted to the data using Eq. (\ref{eq.prec}) from which $\bar{J}$ can be determined.
c) Ratio of the spin-dependent optical dipole force and off-resonant scattering rate, $F_0/\Gamma$, for different beam angles with (red) and without (black) additional EIT cooling. The solid lines show a theoretical calculation as in Fig. \ref{fig.F0_CAD}e. The uncertainty in the data is dominated by the measurement of $\Gamma$, which is shown in the right inset, while the data for $F_0$ is shown in the left inset. We assume a common, systematic uncertainty of $\Delta \theta_\mathrm{ODF}=\qty{0.25}{\degree}$ due the initial alignment.}
\label{fig.F0_meas_new}
\end{figure}

In the following, we characterize the change in the effective spin-spin coupling and optical dipole force magnitude as a function of the ODF beam separation angle, $\theta_\mathrm{ODF}$. We follow the pulse sequence employed in Ref. \cite{Britton2012}, and as shown and described in Fig. \ref{fig.F0_meas_new}a. Here, we vary the initial tipping angle of the spin ensemble, $\theta_1$, which creates a tipping-angle-dependent precession under the application of the ODF. By selecting an ODF frequency detuning, $\delta$, and excitation time, $\tau=2\pi/\delta$, the spin-motional dephasing can be neglected at the end of the ODF interaction sequence, while preserving the effective spin-spin coupling.  The echo sequence allows for the accumulation of spin precession due to the ODF while coherently cancelling other forms of spin precession, e.g. due to magnetic field drifts. The final $(\pi/2)_y$ pulse maps the precession to the spin states.

In this experiment, the applied ODF interaction, $\hat{H}_\mathrm{ODF}$, can be described by an effective mean-field Ising Hamiltonian \cite{Britton2012}, 
\begin{equation}
    \hat{H}_\mathrm{I} 
 = \sum_{i<j}^{N} J_{i,j}  \hat{\sigma}^z_i\hat{\sigma}^z_j,
\end{equation}
where $J_{i,j}$ is the coupling between ion $i$ and $j$. For small positive detunings $\delta$, the Ising interaction becomes independent of the distance between ions, allowing the pairwise spin-spin coupling to be written as \cite{Bohnet_2016}
\begin{equation}
    J_{i,j} \approx \bar{J} = \frac{F_0^2}{4\hbar M\omega_\mathrm{COM}\delta}.
    \label{eq.Jbar}
\end{equation}

Thus, a direct measurement of the uniform coupling parameter $\bar{J}$ allows extraction of the strength of the force $F_0$, which depends on the angle $\theta_{\textrm{ODF}}$ of the ODF beams. In Fig.~\ref{fig.F0_meas_new}b, we illustrate the change in measured $P_{\uparrow}(\theta_{1})$ with and without EIT cooling for different values of $\theta_{\textrm{ODF}}$. As expected, the strength of the induced precession appears to grow with $\theta_{\textrm{ODF}}$ and shows more complex behavior in the presence of effective sub-Doppler cooling. 

From data like those shown in Fig.~\ref{fig.F0_meas_new}b, we first extract $\bar{J}$ via a single-parameter fit to the analytical expression \cite{Britton2012}
\begin{equation}
    P_{\uparrow} = \frac{1}{2}\left( 1+\exp(-\Gamma2\tau)\sin(\theta_1)\sin(2\bar{J}\cos(\theta_1)2\tau)\right),
    \label{eq.prec}
\end{equation}
with $\Gamma$ determined as described before. The dashed lines overlaid with the data show good agreement for all measured values of $\theta_{\textrm{ODF}}$.

Using Eq.~(\ref{eq.Jbar}) and the measurements described above, we can extract the force, $F_0$, for a particular ODF beam angle and cooling method. Since the force, $F_{0}$, is independent of laser detuning, $\delta$, we perform a weighted average of data over multiple values of $\delta$ to mitigate systematic errors that may arise from measuring only at a specific detunings.

In Fig.~\ref{fig.F0_meas_new}c, we show the extracted value $F_{0}/\Gamma$ as a function of $\theta_{\textrm{ODF}}$ ranging between approximately $14\degree$ to $28\degree$. The off-resonant scattering rate, $\Gamma$, is determined independently for each value of $\theta_{\textrm{ODF}}$ and is approximately constant, see the lower right inset of Fig.~\ref{fig.F0_meas_new}c. In addition, the dependence of $F_{0}(\theta_{\textrm{ODF}})$ is shown in the lower left inset. The data agrees well with theoretical calculations, as also shown in Fig.~\ref{fig.F0_CAD}e.

In the Doppler-cooled case, the ratio $F_{0}/\Gamma$ increases only marginally due to the strong decrease of the Debye-Waller factor, compensating the linear increase in force with $\delta k$, see Eq. (\ref{eq.F0}). By contrast, in the presence of EIT cooling, the reduced root-mean-square axial extent of the center-of-mass mode, $z_{\mathrm{COM}} \propto \sqrt{(2\bar{n}+1)}$, increases the Debye-Waller factor as described in Eq. (\ref{eq.F0}). In such a circumstance, the increase in $\delta k$ dominates the measured dynamics and the ratio $F_{0}/\Gamma$ increases quasi-linearly with $\theta_{\textrm{ODF}}$. This results in about a factor two increase in $F_{0}/\Gamma$ between $14\degree$ and $28\degree$, and with and without additional EIT cooling at $28\degree$.

\subsection{Beam stability}
\label{subsec.StabExp}

As a final characterization experiment, we probe the temporal stability of the in-bore optomechanical setup introduced in this work.  The ODF difference wave vector must be aligned parallel to the crystal rotation axis, $\delta \vec{k} \parallel \vec{z}$, to couple only to the axial ion-crystal modes. Misalignment leads to a $(x,y)$-position-dependent ODF, which can be probed by tuning the ODF frequency close to the crystal rotation frequency, $\omega_r$, which results in a spin-dephasing signal similar to the one in Fig. \ref{fig.drumhead_modes}c around $\omega_r$.

We perform a measurement in which we use the amplitude of this signal to monitor the change in beam angle over time. To this end, we first calibrate the system by applying different misalignment beam angles using one of the ODF beam rotation positioners and recording the peak population brightness $P_{\uparrow}$ when scanning $\mu$ over the crystal rotation frequency $\omega_r$. We then fix the ODF detuning frequency at this peak value, $\delta=\mu-\omega_r\approx\pi/\tau$, and measure $P_{\uparrow}$ over an extended duration. The resulting change in population can then be converted into a bound on the angular variation, $\Delta \theta_\mathrm{ODF}$, using the calibration. 

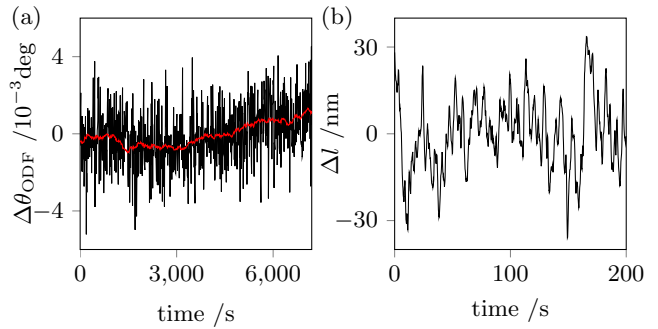
\begin{figure}[h!]%
\centering
\begin{tikzpicture}
\begin{axis}[%
    width = 0.26\textwidth,
    height = 0.26\textwidth,
    name = ODFbeamanglechanges,
    xlabel={time /s},
    ylabel={$\Delta \theta_\mathrm{ODF}$ /$10^{-3}$deg},
    ylabel shift = -8 pt,
    xmin=0, xmax=7200,
    ymin=-6, ymax=6,
    xtick={0,3000,6000},
    ytick={-4,0,4},
    major tick length=\MajorTickLength,
    legend pos=north east,
    legend cell align=left,
    xtick pos=bottom, ytick pos=left,
    xtick align=outside, ytick align=outside,
]
\addplot+[black, sharp plot, no markers] 
table[x=x,y=y1] 
{ODF_beam_angle_stability_data_2.txt};
\addplot+[red, sharp plot, no markers] 
table[x=x,y=y1] 
{ODF_beam_angle_stability_data_1.txt};
\end{axis}%
\node[right] at (ODFbeamanglechanges.left of north west) {(a)};

\begin{axis}[%
    name=pathlengthdifference,
    width = 0.26\textwidth,
    height = 0.26\textwidth,
    xlabel={time /s},
    ylabel={$\Delta l$ /nm},
    ylabel shift = -10pt,
    xmin=0, xmax=200,
    ymin=-40, ymax=40,
    xtick={0, 100, 200},
    ytick={-30, 0, 30},
    major tick length=\MajorTickLength,
    legend pos=north west,
    legend cell align=left,
    xtick pos=bottom, ytick pos=left,
    xtick align=outside, ytick align=outside,
    at = (ODFbeamanglechanges.right of south east), anchor=left of south west,
]
\addplot+[black, sharp plot, no markers] 
table[x=x,y=y1] 
{ODF_beam_phase_stability_data.txt};
\end{axis}%
\node[right] at (pathlengthdifference.left of north west) {(b)};
\end{tikzpicture}%
\caption{ODF beam stability measurements. a) ODF difference wave vector stability $\Delta \theta_\mathrm{ODF}$ determined via coupling to the radial motion. The black curves are individual measurements, and the red curve shows a moving average. The slow drift is on the order of $0.002\,\degree/\mathrm{h}$. b) Path length difference $\Delta l$ of the two ODF beams, measured with an interferometer right before the entrance to the magnet bore.}%
\label{fig.ODF_beam_stab}
\end{figure}

The result is shown in Fig. \ref{fig.ODF_beam_stab}a. We observe a drift of about $\qty{0.002}{\degree/\hour}$. As an example, for a $\qty{300}{\micro\meter}$-diameter ion crystal and an ODF beam separation angle of $28^\circ$ as used above, having a beat note wavelength of about $\qty{647}{\nano\meter}$, a misalignment of $\qty{0.002}{\degree}$ results in an ODF phase difference of $\qty{2.9}{\degree}$ from the central ion to the outermost ions.  This difference is small relative to other sources of experimental uncertainty.

Furthermore, we evaluate the relative optical phase stability of both ODF beams. As shown in Fig. \ref{fig:trap_optics}, both ODF beams are generated by splitting a single beam via a polarising beamsplitter, and then travel along spatially separated paths before they are combined again by a knife-edge prism and travel close-neighbouring beam paths until the ion crystal.  While the beams are spatially separated, they are susceptible to interferometric errors due to environmental fluctuations. When travelling through the periscope, we pick off a part of both beams and combine them interferometrically on a photodiode. The path-length difference over a measurement time of several minutes is shown in Fig. \ref{fig.ODF_beam_stab}b. Slow fluctuations of $\ll \qty{1}{\hertz}$ can have amplitudes of around $\Delta l \approx \qty{20}{\nano\meter}$, while higher-frequency fluctuations have a much smaller amplitude of $<\qty{5}{\nano\meter}$. The root-mean-square of the differential beam-path fluctuation is $\Delta l_\mathrm{rms}\approx\qty{12}{\nano\meter}$, corresponding to about $\qty{7}{\degree}$ phase fluctuation of the ODF beatnote at a separation angle of $28^\circ$. We note that it is possible to compensate for this path length fluctuation using closed-loop feedback on a piezo-mounted mirror in one of the ODF beamlines (not shown in Fig. \ref{fig:trap_optics}).

\section{Summary and Outlook}
\label{sec.sum}

In this work, we have presented an optomechanical system that enables precise alignment and in-situ tuneable spin-spin interaction strength in 2D ion crystals confined in a Penning trap. This system uses piezo-actuated positioners to manipulate laser beams in the tightly confined space of the bore of a superconducting magnet, facilitating beam alignment precision, reproducibility and ease of use. We achieve an increase of a factor of two in coherent interaction strength by increasing the beam separation angle between two Raman beams that create a spin-dependent optical dipole force, and by applying EIT near-ground state cooling via another pair of beams. At the same time, the major source of decoherence, the off-resonant scattering due to these beams, is not affected, ultimately increasing the effective coherent-to-incoherent coupling strength of spin-motion and spin-spin interactions.  The ODF beam alignment stability was determined to be on the order of $\qty{0.002}{\degree/\hour}$, leading to ODF phase drifts of only a few degrees over a crystal of a few $\qty{100}{\micro\meter}$ diameter, which can be considered small. The differential phase fluctuation in both ODF beam paths due to environmental effects is on the same order of magnitude. Further improvements to the effective coherent to incoherent interaction strength are possible, for example, by using parametric amplification \cite{Affolter2023}.

We demonstrate EIT cooling of 2D ion crystals using two phase-coherent, $\qty{55}{\giga\hertz}$-separated laser beams generated by modulation of the same laser. We achieve near-ground state cooling down to $\bar{n}\approx 1.3$ in $\qty{850}{\micro\second}$, currently limited by the laser power. A future power upgrade should allow further cooling in reduced time \cite{Jordan2019}, improving experimental efficiency.

In general, these developments will play an important role in future quantum simulation and sensing applications using 2D and potentially 3D ion crystals \cite{Hawaldar2023}.  In particular, this combination of technical advances will enable access to new experimental regimes that require precise control of the tilting of the ODF wavefront to generate a coupling to the radial motion \cite{Shankar2022a}.\\

\section*{Acknowledgements}

The project was supported by the Australian Research Council Centre of Excellence for Engineered Quantum Systems (CE170100009) and a private grant from H. and A. Harley. R.N.W. acknowledges support from the Australian Research Council under the Discovery Early Career Researcher Award scheme (DE190101137). J.H.P and J.Y.Z.J. acknowledge support through the Sydney Quantum Academy and an Australian Government Research Training Program (RTP) Scholarship.

\bibliography{references}

\end{document}